\documentclass[prb,twocolumn,nofootinbib,superscriptaddress]{revtex4-1} 
\usepackage{amsmath}
\usepackage{bbm}
\usepackage{graphicx}
\usepackage{color}
\usepackage{amssymb}
\usepackage[normalem]{ulem}
\usepackage[colorlinks=true,linkcolor=blue,citecolor=blue,urlcolor=blue]{hyperref}
\usepackage{siunitx}

\newcommand{\bea}{\begin{eqnarray}}
\newcommand{\eea}{\end{eqnarray}}
\newcommand{\be}{\begin{equation}}
\newcommand{\ee}{\end{equation}}
\newcommand{\benn}{\begin{equation*}}
\newcommand{\eenn}{\end{equation*}}

\newcommand{\kBT}{k_\text{B}T}

\newcommand{\e}{\mathrm{e}}

\newcommand{\rme}{\text{e}}
\newcommand{\rmo}{\text{o}}
\newcommand{\CAR}{\text{CPS}}
\newcommand{\CPS}{\text{CPS}}
\newcommand{\EC}{\text{EC}}
\newcommand{\COP}{\text{COP}}
\newcommand{\rmL}{\mathrm{L}}
\newcommand{\rmR}{\mathrm{R}}
\definecolor{light-gray}{gray}{0.9}
\definecolor{gris}{gray}{0.5}

\begin{document}

\title{
\vspace{-2.25cm}
\textnormal{\small  PHYS. REV. B {\bf 98}, 241414(R) (2018)}\\
\vspace*{-0.2cm}
\rule[0.1cm]{18cm}{0.02cm}\\
\vspace*{0.2cm}
Cooling by Cooper pair splitting}
\author{Rafael S\'anchez}
\affiliation{Departamento de F\'isica Te\'orica de la Materia Condensada, Condensed Matter Physics Center (IFIMAC) and Insituto Nicol\'as Cabrera, Universidad Aut\'onoma de Madrid, 28049 Madrid, Spain}
\author{Pablo Burset}
\affiliation{Department of Applied Physics, Aalto University, 00076 Aalto, Finland}
\author{Alfredo Levy Yeyati}
\affiliation{Departamento de F\'isica Te\'orica de la Materia Condensada, Condensed Matter Physics Center (IFIMAC) and Insituto Nicol\'as Cabrera, Universidad Aut\'onoma de Madrid, 28049 Madrid, Spain}
\date{\today}

\begin{abstract}
The electrons forming a Cooper pair in a superconductor can be spatially separated preserving their spin entanglement by means of quantum dots coupled to both the superconductor and independent normal leads. 
We investigate the thermoelectric properties of such a Cooper pair splitter and demonstrate that cooling of a reservoir is an indication of non-local correlations induced by the entangled electron pairs. 
Moreover, we show that the device can be operated as a non-local thermoelectric heat engine. Both as a refrigerator and as a heat engine, the Cooper pair splitter reaches efficiencies close to the thermodynamic bounds. 
As such, our work introduces an experimentally accessible heat engine and a refrigerator driven by entangled electron pairs in which the role of quantum correlations can be tested. 
\end{abstract}
\maketitle

{\it Introduction.---} Hybrid nanostructures with superconducting or normal electrodes connected by quantum wires
provide a unique playground to test the interplay between transport and electron correlations \cite{hybrid-review}. Among these types of devices, Cooper pair splitters have received special attention due to their potential use as a source of non-locally entangled electron pairs \cite{recher,martin,feinberg}. 
A typical device consists of a central superconducting lead coupled to two normal ones through quantum dots, a geometry
which has been successfully implemented using semiconducting nanowires \cite{Hofstetter,Heiblum,Fullop,Tarucha}, carbon nanotubes
\cite{Herrmann,Schindele} or graphene \cite{Hakonen}. In this setup the Coulomb repulsion in the quantum dots forces the incoming electron pairs from the superconductor to separate into different normal electrodes, while local Andreev processes in which the two electrons from the pair are transferred to the same normal lead are strongly suppressed \cite{recher}. The Cooper pair splitting (CPS) process can be viewed as the time reverse of a crossed Andreev reflection in which an incoming electron from a given normal lead is reflected as a hole in the opposite lead \cite{Beckmann,Morpurgo}. Positive correlation between the currents through the two normal contacts has been presented as a signature of the splitting process~\cite{Hofstetter,Herrmann,Schindele}. 

Conventional transport measurements, including noise correlations~\cite{Chtchelkatchev,samuelsson_orbital_2003,bignon_current_2004}, could allow for a complete characterization of the underlying microscopic processes~\cite{Burset} and, eventually, entanglement detection~\cite{Braunecker}. However, the thermal properties of Cooper pair splitters provide another perspective to the problem which still needs to be explored, with the exception of thermoelectrically induced CPS~\cite{cao_thermoelectric_2015}. 
While these setups could have their own interest as multiterminal thermoelectric devices~\cite{machon_nonlocal_2013}, in which the separation of heat and charge currents could be achieved \cite{review,Mazza2015}, we rather focus on the thermoelectric properties which may appear as signatures of the presence of Cooper pair splitting. 
We show that, by virtue of these quantum correlated processes, heat can flow from a cold normal lead to a hotter one. 
This mechanism is different from that used in systems based on Peltier cooling~\cite{giazotto_opportunities_2006} or heat drag~\cite{sinis}, and from absorption refrigerators~\cite{mari_cooling_2012,cleuren_cooling_2012,chen_quantum_2012,entin_enhanced_2015} or heat pumps~\cite{liliana,miguel}, since no voltage bias is applied between the two leads, and energy is conserved. 

\begin{figure}[b]
\includegraphics[width=\linewidth,clip]{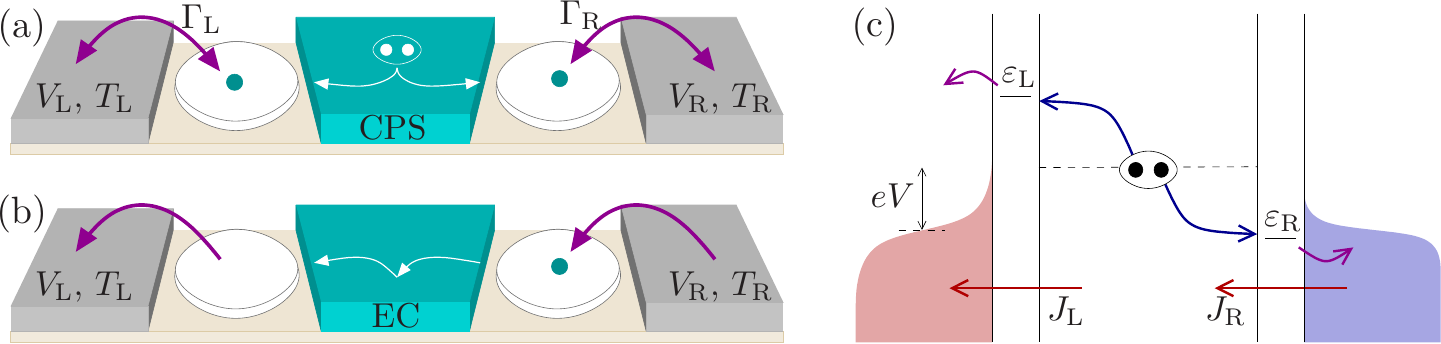}\\
\caption{\label{scheme}\small Quantum dot Cooper pair splitter. (a,b) Sketch of the device: the central superconductor is coupled to two normal terminals (at voltage and temperature $V_l$ and $T_l$, with $l{=}{\rmL,\rmR}$) through two quantum dots with tunneling rates $\Gamma_l$. Transport at subgap voltages is due to (a) CPS and (b) EC processes. (c) For properly tuned dot levels, $\varepsilon_{\rmL,\rmR}$,  and finite $V{=}V_\rmL{=}V_\rmR$, the splitting of a Cooper pair results in heat extraction from one terminal ($J_\rmR{>}0$),  against a thermal gradient ($T_\rmL{>}T_\rmR$).}
\end{figure}

The basic idea is illustrated in Fig.~\ref{scheme}. There are two elementary processes which contribute to transport when bias voltage and temperatures are much smaller than the dot charging energies and the superconducting gap: the already described CPS processes [Fig.~\ref{scheme}(a)] and elastic cotunneling (EC) processes in which an electron is transferred between the normal leads without changing the charge of the superconductor [Fig.~\ref{scheme}(b)]. Let us consider the situation depicted in Fig.~\ref{scheme}(c) in which the left electrode temperature is larger than the right one ($T_\rmL{>}T_\rmR$) but their chemical potentials are equal and below the chemical potential of the superconductor ($\mu_{\rm S}{>}\mu_\rmL{=}\mu_\rmR$). Then, if the dot levels $\varepsilon_{\rmL,\rmR}$ are tuned to be opposite and $\varepsilon_\rmR{<}\mu_\rmR$, the CPS processes inject electrons below the Fermi energy in the right lead (and over it in the left one), hence contributing to cooling the right and heating the left lead. 
As we discuss in detail below, although EC tends to destroy this effect, it remains robust for a broad range of parameters. Therefore, our proposal should be feasible using the current generation of Cooper pair splitting devices. 

{\it Model.---} We consider a quantum dot based Cooper pair splitter: two quantum dots serve as the links between a superconductor and two normal leads, as sketched in Fig.~\ref{scheme}. We assume that the superconducting gap $\Delta$ and local Coulomb interactions $U_0$ are large so that double occupancy of a single dot can be neglected (typical experiments~\cite{Hofstetter,Heiblum,Fullop,Schindele} are performed at $T\sim20-50$~\si{mK}, with $\Delta\sim100\!-\!200$~\si{\mu eV} and $U_0>1$~\si{meV}).
Thus our model Hamiltonian of the system can be written as
\be
\hat H=\sum_{l{=}\rmL,\rmR} \varepsilon_l\hat n_l+U\hat n_\rmL\hat n_\rmR+\hat H_\tau+\hat H_{\rm tun}+\hat H_{\rm leads},
\ee
where $\hat n_l{=}\sum_\sigma d_{l\sigma}^\dagger d_{l\sigma}$ is the occupation operator of quantum dot $l$, with $d_{l\sigma}^\dagger$ the dot electron creation operator with spin $\sigma$, $U$ is the interdot Coulomb interaction, $\hat H_{\rm leads}$ and $\hat H_{\rm tun}$ describe the normal leads and the lead-dot coupling, and
\be
\hat H_\tau{=}{-}\sum_\sigma\tau_\EC \hat d_{\rmL\sigma}^\dagger\hat d_{\rmR\sigma}{-}\frac{\tau_\CAR}{\sqrt2}{\left(\hat d_{\rmL\uparrow}^\dagger\hat d_{\rmR\downarrow}^\dagger{-}\hat d_{\rmL\downarrow}^\dagger\hat d_{\rmR\uparrow}^\dagger\right)}{+}{\rm H.c.}
\ee
is the effective Hamiltonian resulting from integrating out the superconducting lead~\cite{feinberg,eldridge,Braggio,bjorn,Trocha,amitai_nonlocal_2016,robert,
Trocha2,walldorf}. It describes EC and CPS non-local processes mediated by the superconductor with amplitudes $\tau_{\EC}$ and $\tau_\CPS$, respectively.

The CPS and EC terms hybridize the even-parity, $|00\rangle$ and $|{\rm S}\rangle=\frac{1}{\sqrt{2}}(|{\uparrow}{\downarrow}\rangle-|{\downarrow}{\uparrow}\rangle)$, and the odd-parity states, $|\sigma0\rangle$ and $|0\sigma\rangle$, leading to the double dot states,
\begin{align}
\label{even}
|\e,\pm\rangle&=N_{\e\pm}^{-1}\left(\tau_\CAR|00\rangle-E_{\e\pm}|{\rm S}\rangle\right), \\
\label{odd}
|{\rmo,\sigma,\pm}\rangle&=N_{\rmo\pm}^{-1}\left(\tau_\EC|\sigma0\rangle-E_{\rmo\pm}|0\sigma\rangle\right),
\end{align}
with $N_{\alpha\pm}$ fixed by normalization, and (eigen)energies $E_{\e\pm}=E_\pm(0,\varepsilon_\rmL{+}\varepsilon_\rmR{+}U,\tau_\CAR)$ and $E_{\rmo\pm}=E_\pm(\varepsilon_\rmL,\varepsilon_\rmR,\tau_\EC)$, where $E_\pm(x,y,\tau)=\left[x{+}y{\pm}\sqrt{(x{-}y)^2{+}4|\tau|^2}\right]/2$. 
On top of these states, the Hilbert space includes the triplet states $|{\rm T}_0\rangle=\frac{1}{\sqrt{2}}(|{\uparrow}{\downarrow}\rangle{+}|{\downarrow}{\uparrow}\rangle)$, $|{\rm T}_+\rangle=|{\uparrow}{\uparrow}\rangle$, and $|{\rm T}_-\rangle=|{\downarrow}{\downarrow}\rangle$, all of them with energy $E_{\rm T}=\varepsilon_\rmL+\varepsilon_\rmR+U$.

In the sequential tunneling regime, the coupling to the leads can be incorporated to lowest order in perturbation theory. 
This is characterized by tunneling rates $\Gamma_{\rmL/\rmR}\ll\tau_{\CPS/\EC},\kBT_l$, which introduce transition rates between quantum dot states ($i{\rightarrow}j$) given by
\begin{align}
\Gamma_{ji}^{l+}&=\Gamma_l|\langle j|d_{l\sigma}^\dagger|i\rangle|^2f_l(E_{j}{-}E_i-eV_l), \\
\Gamma_{ji}^{l-}&=\Gamma_l|\langle j|d_{l\sigma}|i\rangle|^2[1-f_l(E_{i}{-}E_j-eV_l)],
\end{align}
where $f_l(E)=\left[1+\exp(E/\kBT_l)\right]^{-1}$.
Note that since the even parity states do not have a well defined number of particles, transitions between the even and odd states can be due to either electron or hole tunneling events. 

The charge and heat currents in the normal leads are written as $I_l={\cal I}_l[c_{l\pm}]$, and $J_l={\cal I}_l[h_{l\pm}]$, where
\begin{align}
{\cal I}_l[\lambda]&=\sum_{j,i\neq j,\alpha}\lambda_{l\alpha}\Gamma_{ji}^{l\alpha}p_i,
\end{align}
$c_{l\pm}{=}{\pm}e$, and $h_{l\pm}{=}E_j{-}E_i{\pm}eV_l$. 
They depend on the stationary occupation of the different states, $p_i$, which is obtained from the master equation
\be
\dot p_i=\sum_{j,l,k}\left(\Gamma_{ij}^{lk}p_j-\Gamma_{ji}^{lk}p_i\right).
\ee
Note that within our approximations, the evolution of the non-diagonal elements of the density matrix is decoupled from the populations $p_i$ in the eigenstate basis.
A crucial aspect in this device is that the coupling to the superconductor injects particles but conserves energy in the normal subsystem, i.e., $I_\rmL+I_\rmR\neq0$ and
\be
\label{eq:energycons}
\sum_{l=\rmL,\rmR}(I_lV_l+J_l)=0.
\ee
Only when $\tau_\CPS=0$ do we have $I_\rmL+I_\rmR=0$.

\begin{figure}[t]
\includegraphics[width=0.95\linewidth,clip]{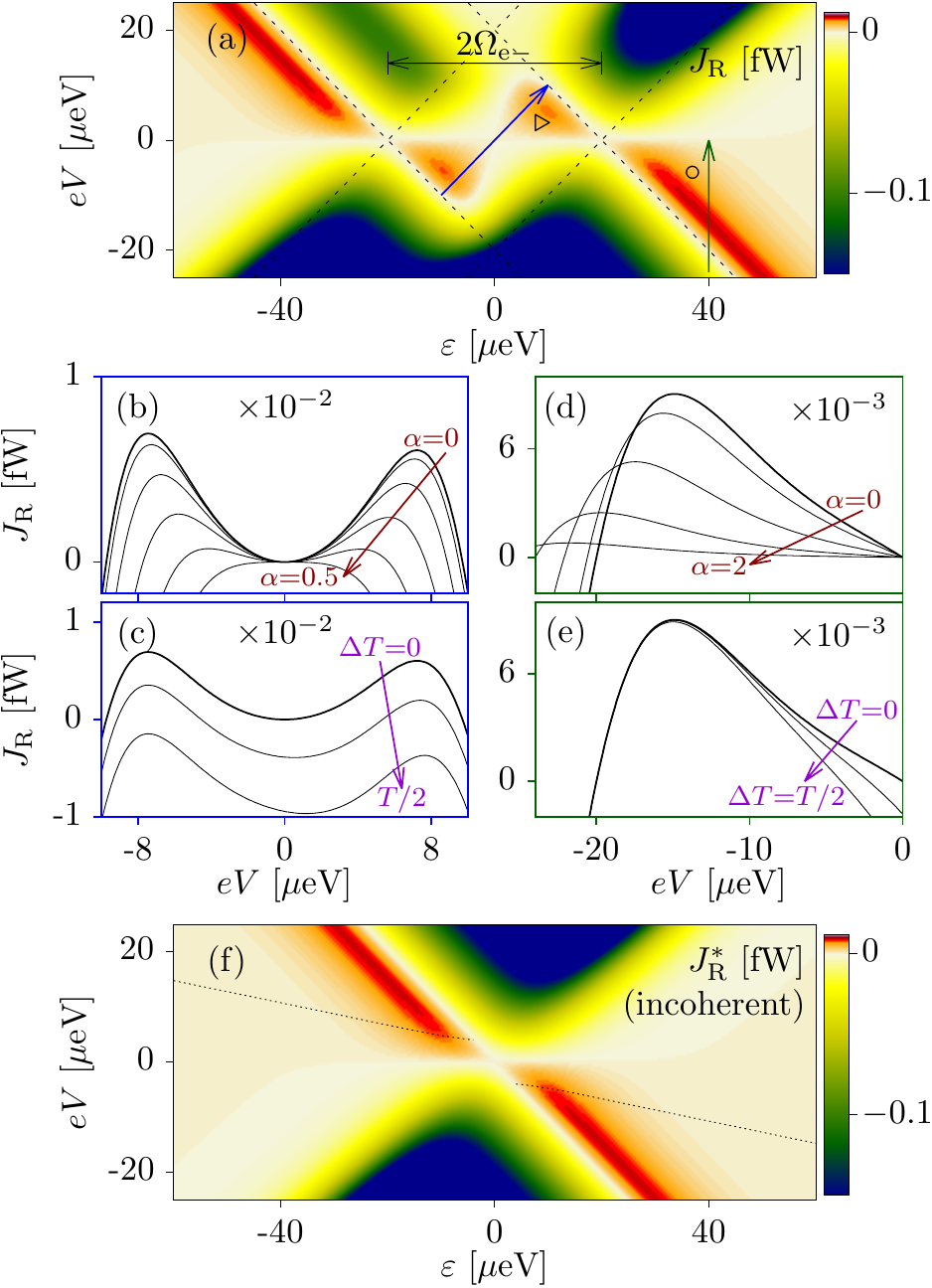}
\caption{\label{cooling}\small CPS cooling. (a) Heat current injected from the right terminal as a function of the quantum dot levels $\varepsilon\!=\!\varepsilon_\rmL\!=\!-\varepsilon_\rmR$ and applied bias voltage $V\!=\!V_\rmL\!=\!V_\rmR$. Cooling occurs in the regions with $J_\rmR\!>\!0$. Dashed lines indicate the relevant transitions between the quantum dot states. Cuts along the blue and green arrows are plotted in (b-e), compared to the cases with an increased contribution of the elastic cotunneling $\alpha=\tau_\EC/\tau_\CAR$ (b,d) or temperature gradient $\Delta T=T_\rmR-T_\rmL$ (c,e). Parameters: $U=0$, $T=23.2$~\si{mK}, $\tau_\CAR=20$~\si{\mu eV}, $\hbar\Gamma=2$~\si{\mu eV}, $\tau_\EC=0$, and $\Delta T=0$ except when explicitly stated. (f) Heat current for the incoherent setup with two quantum dots coupled to a thermal reservoir whose temperature $T_{\rm C}$ adapts to fulfill Eq.~\eqref{eq:energycons}, with $\gamma=\Gamma/3$. Dotted lines mark the configurations with cooling at a fixed $T_{\rm C}=5T/4$. 
}
\end{figure}

{\it CPS cooling.---} We now demonstrate the cooling effect of CPS processes with a simple configuration that can be described analytically. We consider the case with antisymmetric energy levels $\varepsilon\equiv\varepsilon_\rmL=-\varepsilon_\rmR$ where CPS is most effective. For simplicity, we first restrict to the situation with $\tau_\EC=0$, where the odd states do not hybridize. In the region with $\varepsilon\gg|\tau_\CPS|\gg\kBT_\rmL\ge\kBT_\rmR$ and $eV_\rmL=eV_\rmR=eV<0$, transport is dominated by the states $|\e,-\rangle$ and $|0,\sigma\rangle$. This is the configuration sketched in Fig.~\ref{scheme}(c) and marked by a green arrow in Fig.~\ref{cooling}(a). The opposite configuration with $\varepsilon<0$, $eV>0$ can be treated equivalently. Only two transitions (and their reversed) are hence relevant: starting from the doubly occupied state, one electron can tunnel to the left lead with a rate $(\Gamma_\rmL/4)[1-f_\rmL(x_+)]$, and the remaining one tunnels to the right one with rate $(\Gamma_\rmR/2)[1-f_\rmR(x_-)]$. For the reversed transitions, one has to replace $1{-}f(x_\pm){\rightarrow}f(x_\pm)$. Here $x_\pm=\Omega_{\e-}{-}\varepsilon{\pm}eV$ and  $2\Omega_{\e-}\equiv\sqrt{U^ 2{+}4|\tau_\CPS|^ 2}-U$. Then, the initial state is restored by the splitting of a Cooper pair. 
Solving the master equation for the occupation of these two states, one arrives at the heat currents
\be
J_\rmR=x_-\frac{\Gamma_\rmL\Gamma_\rmR}{2\Gamma_\rmL{+}4\Gamma_\rmR}[f_\rmR(x_-)-f_\rmL(x_+)]
\ee
and $J_\rmL=-(x_+/x_-)J_\rmR$. 
Note that transitions in the left and right barriers occur at different energies split by $x_+-x_-=2eV$. 
For $\Omega_{\e-}{-}\varepsilon<eV<\varepsilon{-}\Omega_{\e-}$, heat is extracted from the right terminal ($J_\rmR{>}0$) and is absorbed by the left one ($J_\rmL{<}0$). 
Hence, if $T_\rmL{>}T_\rmR$, heat flows from the cold to the hot terminal in this regime. 
Such a CPS-assisted cooling effect is different from other quantum dot based refrigerators that exploit a voltage gradient between the two terminals or the energy exchange with an additional thermal bath~\cite{entin_enhanced_2015,thermal-long}. 

One can also easily check that the charge currents fulfill $I_\rmL=I_\rmR=(e/x_-)J_\rmR$. They have the same sign, as expected for CPS processes. The proportionality between $I_l$ and $J_l$ achieves a tight-coupling limit, at which thermoelectric processes attain high efficiencies~\cite{benenti_fundamental_2017}, as will be discussed later. 

The behaviour of the heat current $J_\rmR$ as a function of the different parameters is illustrated in Figs.~\ref{cooling}(a)--(e). Let us concentrate on the case with $\varepsilon>0$ (the opposite case can be understood with similar arguments). We can distinguish two regions where cooling occurs, [marked by $\triangleright$ and $\circ$ in Fig.~\ref{cooling}(a)] depending on the relative value of $|\varepsilon|$ and $\Omega_{\e-}$. The case for $\varepsilon>\Omega_{\e-}$, $eV<0$ ($\triangleright$) can be understood in terms of the analytical model presented above. 
In this regime, the energy of the odd states dominates the heat currents, so CPS cooling is robust against the contribution of EC processes. It also survives for relatively high temperature gradients, cf. Figs.~\ref{cooling}(d) and (e). Although the magnitude of the cooling power can be enhanced by increasing the base temperature, we have chosen to illustrate the effect using conservative parameters that warrant the validity of our simplified description.

The case for the region $\varepsilon<\Omega_{\e-}$, $eV>0$ ($\circ$) is different. There the dot chemical potentials are dominated by the coherent hybridization due to $\tau_\CPS$, see Eq.~\eqref{even}. The positive $eV$ induces the absorption of two electrons from the left and right leads and their recombination as a Cooper pair, mainly through the sequence $|\rme,-\rangle\rightarrow|0\sigma\rangle\rightarrow|\rme,-\rangle$. In the first transition, an amount of heat $x_-$ is extracted from the right terminal. Then, $-x_+$ is absorbed by the left one as heat. Hence, cooling is driven by a purely quantum mechanical process in this regime. Contrary to the previous case, this regime is more sensitive to EC and $J_\rmR$ becomes negative as $\alpha$ exceeds 0.5, cf. Figs.~\ref{cooling}(b) and (c). 

Note also that for $eV>0$ and $\varepsilon\approx0$ heat currents are suppressed due to the occupation of the triplet states, which leads to a spin blockade~\cite{eldridge} of heat currents. 

{\it Comparison to a incoherent setup.---} 
To emphasize the essential role of quantum correlations, it is useful to compare our case with a similar system of two independent quantum dots coupled to a fictitious third normal electronic reservoir, C, whose temperature $T_{\rm C}$ adapts to the condition of no energy injection into the system~\cite{meair_local_2014}, cf. Eq.~\eqref{eq:energycons}. Such a geometry has been previously investigated for cryogenic purposes~\cite{edwards,prance,swiss}. 
For simplicity, let us assume $U{=}0$ and $\varepsilon=\varepsilon_\rmL=-\varepsilon_\rmR$, and that all tunneling rates are equal and given by $\gamma$. Then, cooling occurs in the regions with $eV<-\varepsilon$ where $J_\rmR^*=(\varepsilon+eV)\gamma[f_\rmR(\varepsilon+eV)-f_\rmL(\varepsilon-eV)]/4$, see Fig.~\ref{cooling}(f).
However, cooling in this setup is only possible due to an additional condition, namely that C must become the hottest reservoir: $T_{\rm C}\ge T_\rmL\ge T_\rmR$. Furthermore, $T_{\rm C}$ depends on the voltage and level configuration to satisfy Eq.~\eqref{eq:energycons}. 
In a realistic configuration with a well defined $T_{\rm C}$, cooling under these conditions is only possible for particular configurations --an example is marked with dotted lines in Fig.~\ref{cooling}(d). 
In contrast, in the coherent case CPS warrants energy conservation for every configuration. Note also by comparing Figs.~\ref{cooling}(a) and (d), that in the coherent case cooling is achieved for regions with $\varepsilon$ and $eV$ having the same sign (for $|\varepsilon|<\Omega_{\e-}$, region $\triangleright$), something that is not possible in the system with no quantum correlations. The role of spin correlations is also evident in the absence of a spin blockade in the incoherent case, leading to the electron-hole symmetric Fig.~\ref{cooling}(f).


{\it Heat engine.---} As the two dots are not directly coupled, the superconductor also mediates a non-local thermoelectric effect. A charge current flows in one terminal by increasing the temperature of the other one~\cite{cao_thermoelectric_2015}. 
The non-conservation of charge due to CPS strongly influences the system response. For instance, if $\tau_\EC{=}0$, the same number of electrons are injected in both leads, so no thermovoltage will develop between them. 
We study two relevant configurations to operate the Cooper pair splitter as a heat engine, assuming terminal L is hot, $T_\rmL=T_\rmR+\Delta T$: (i) short circuit, where L is grounded ($V_\rmL{=}0$), and (ii) open circuit, where L acts as a voltage probe, i.e., a floating $V_\rmL$ develops such that it injects heat but not charge ($I_\rmL=0$). 

\begin{figure}[t]
\includegraphics[width=\linewidth,clip]{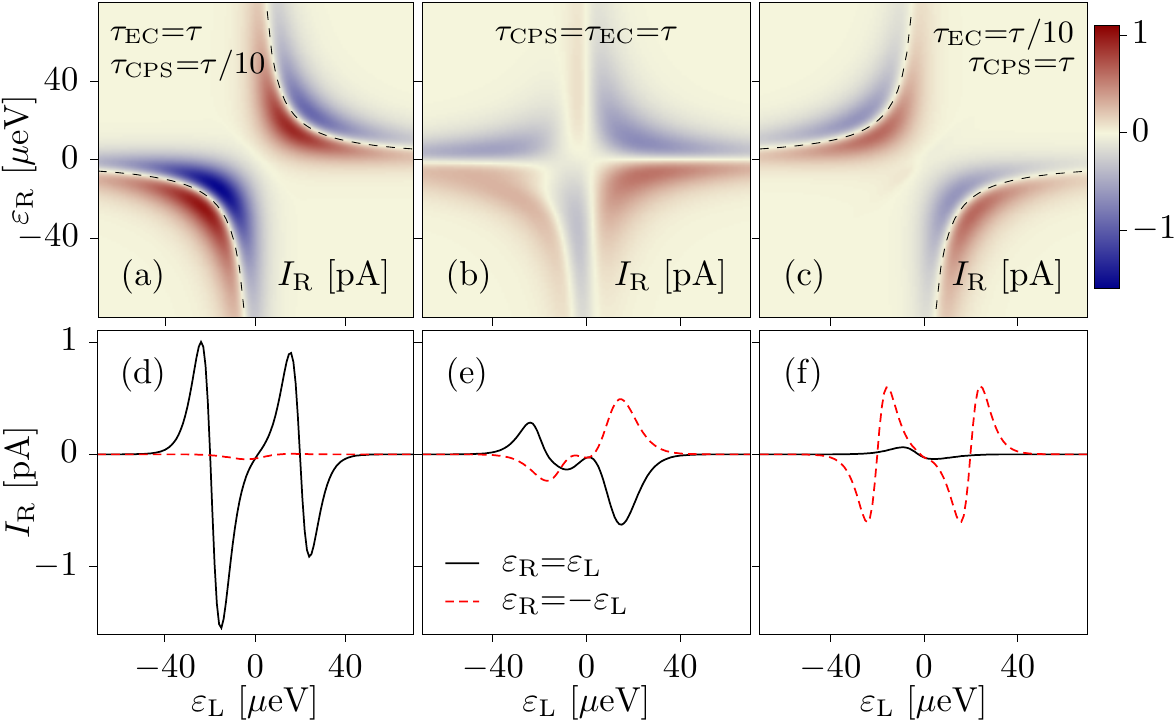}
\caption{\label{curr}\small (a-c) Thermoelectric current in terminal R as a function of the quantum dot levels, for different relative contributions of the EC and CPS processes. Black-dashed lines mark the crossing of the relevant eigenstates involved in the response. Panels (d-f) show cuts of the ones above them along the symmetric ($\varepsilon_\rmL{=}\varepsilon_\rmR$) and antisymmetric ($\varepsilon_\rmL{=}{-}\varepsilon_\rmR$) conditions. Same parameters as in Fig.~\ref{scheme}(c), except for $V=0$. }
\end{figure}

Let us first consider the short circuit configuration (i). Transport strongly depends on whether CPS or EC processes dominate, as it is shown in Fig.~\ref{curr}. The case with dominant EC exhibits a typical double quantum dot response~\cite{holger-dqd}, where transport in the center of the stability diagram is dominated by interdot tunneling, cf. Fig.~\ref{curr}(a). The current shows a characteristic double oscillation around the points where the number of particles 
of the double dot changes
as $\varepsilon_\rmL=\varepsilon_\rmR$ increases, see Fig.~\ref{curr}(d).

Differently, the case with dominant CPS exhibits an inverted stability diagram, where the largest oscillations occur along the condition $\varepsilon_\rmL=-\varepsilon_\rmR$, see Figs.~\ref{curr}(c,f). 
For this condition, the two current oscillations result from transitions between $|\e,-\rangle$ and the corresponding odd state $|\rmo,\sigma,\pm\rangle\approx|\sigma,0\rangle,|0,\sigma\rangle$.

Available experiments suggest that both CPS and EC contributions might be of the same order~\cite{Hofstetter,Heiblum,Fullop,Herrmann,Schindele,Hakonen}, which gives a mixed thermoelectric effect, cf. Fig.~\ref{curr}(b). This is relevant for the open circuit configuration (ii). As discussed above, if $\tau_\EC{=}0$ or $\tau_\CPS{=}0$ we have $I_\rmR{=}{\pm}I_\rmL$. In both cases the system delivers no power, since we impose $I_\rmL{=}0$ and hence also $I_\rmR{=}0$. Consequently, the optimal configuration for an open circuit engine requires that both EC and CPS contributions are finite and of the same order, see below. 


\begin{figure}[t]
\includegraphics[width=\linewidth,clip]{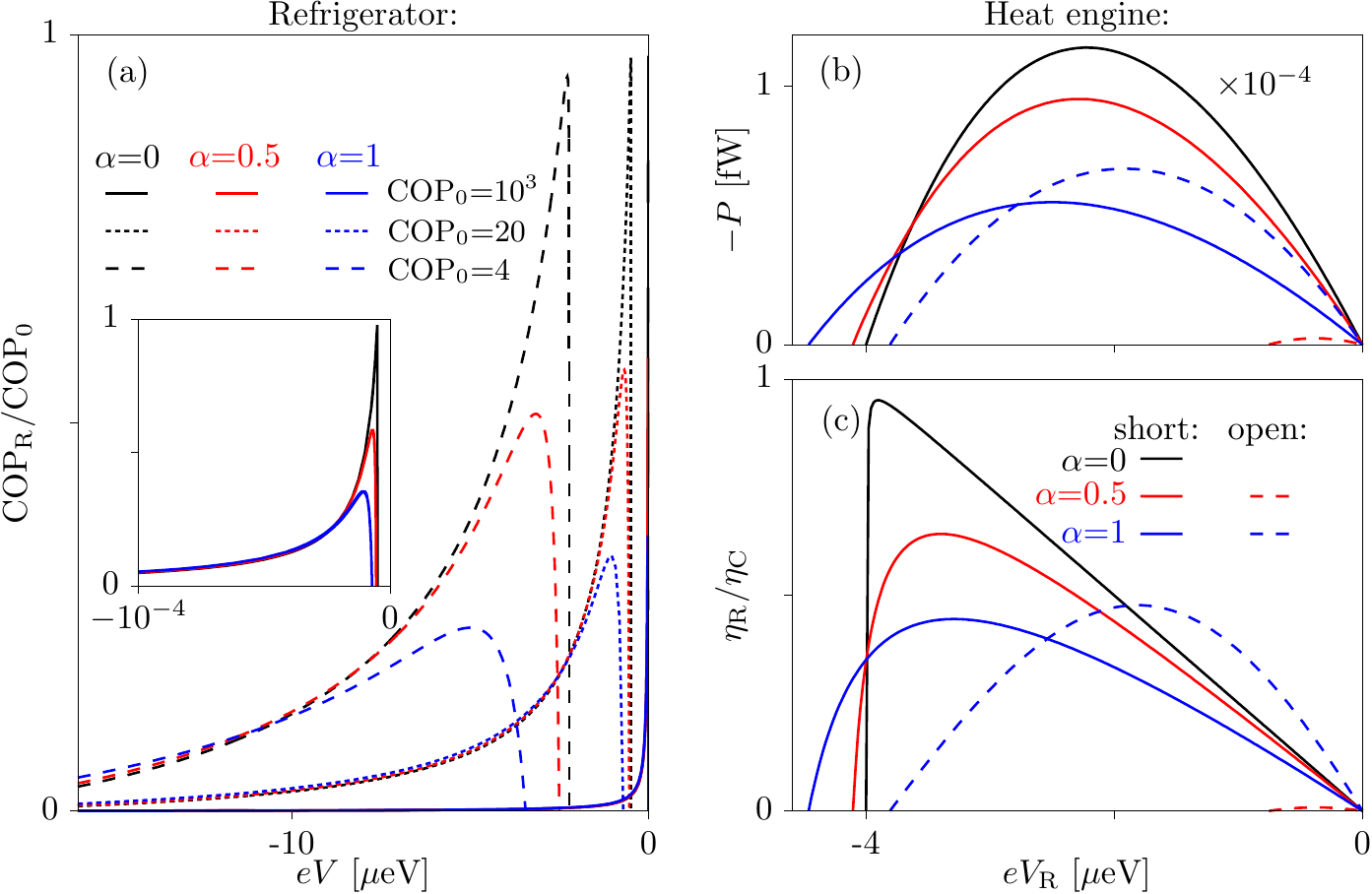}
\caption{\label{eff}\small Performance of the (a) refrigerator and (b,c) heat engine configurations, with different relative contribution of the EC and CPS processes, $\alpha{=}\tau_\EC/\tau_\CPS$. (a) Coefficient of performance at different $\Delta T$ parametrized by the optimal COP$_0$. (b) Generated power and (c) efficiency for short circuit ($V_\rmL{=}0$) and open circuit ($I_\rmL{=}0$) configurations, with $\Delta T{=}T_\rmR/4$. Note that no power is generated in open circuit for the case $\alpha{=}0$. All other parameters as in Fig.~\ref{cooling}(d).
}
\end{figure}

{\it Efficiencies.---} We now discuss the performance of the system as a refrigerator and as a heat engine. 
This is done in terms of the cooling coefficient of performance $\COP={J_\rmR}/{P}$ for the refrigerator, and of the thermoelectric efficiency $\eta={-P}/{J_\rmR}$, for the heat engine, where $P=\sum_lI_lV_l$ is the power. By thermodynamic considerations, one can show that both parameters are bounded by $\COP_0=T_\rmR/\Delta T$ and $\eta_{\rm C}=1-T_\rmR/T_\rmL$, respectively.

For cooling, the COP gets arbitrarily close to the optimal value when only CPS contributes, independently of the temperature gradient, as shown in Fig.~\ref{eff}(a). A finite $\Delta T$ also increases the voltage at which the optimal performance is obtained. Note that COP diverges for $\Delta T{=}0$. The cooling power is robust against the presence of EC and/or a finite temperature gradient, cf. Figs.~\ref{cooling}(d) and (e), but the COP becomes lower, around $0.3\COP_0$, when $\tau_\EC\approx\tau_\CPS$.

For the heat engine, we again distinguish the open and short circuit configurations. In both cases, $I_\rmL V_\rmL=0$, so power is only finite in the right terminal. Short circuiting the left terminal achieves the largest generated power when CPS dominates, cf. Fig.~\ref{eff}(b). In this configuration, the engine furthermore works close to the Carnot bound, as shown in Fig.~\ref{eff}(c). Both the power and the efficiency decrease with increasing $\tau_\EC$, with $\eta{\approx}0.4\eta_{\rm C}$ for $\tau_\EC\approx\tau_\CPS$.


As discussed above, the open circuit gives no power (or very tiny) if only one microscopic process is present. Interestingly though, when both processes have similar contributions, $\tau_\EC{\approx}\tau_\CPS$, the power and the efficiency are better than at short circuit. In this case, the highest efficiency coincides with the maximum power extraction. 


{\it Conclusions.---} We have analyzed the thermoelectric properties of a Cooper pair splitter, identifying configurations where the reversed heat current from a cold to a hot reservoir with the same chemical potential is an indication of correlations mediated by Cooper pair splitting. These microscopic processes also play a key role in the dual operation of the system as a heat engine. 
The Cooper pair splitter is thus an example of a tunable device that utilizes quantum correlations to perform thermodynamic operations~\cite{hofer_quantum_2016,hofer_autonomous_2016,karimi_otto_2016,
marchegiani_self_2016,solinas_microwave_2016,tan_quantum_2017} with high efficiencies.

Finally, let us mention that the predicted cooling power, of the order of fractions of fW for experimentally relevant configurations, is within the achieved resolution in recent measurements of Coulomb blockaded heat currents~\cite{prance,schreider,dutta,sivre}. 

We thank Z. Tan, P. Hakonen, D. Golubev and G. Lesovik for discussions. We acknowledge financial support from the Spanish MINECO via grants FIS2015-74472-JIN (AEI/FEDER/UE), FIS2014-55486P, FIS2017-84860-R, the MAT2016-82015-REDT network and through ``Mar\'\i a de Maeztu" Programme for Units of Excellence in R\&D (MDM-2014-0377), the Ram\'on y Cajal program RYC-2016-20778, and the EU Horizon 2020 research and innovation program under the Marie Sk\l odowska-Curie Grant No. 743884. 

{\it Note added.---} Upon the completion of this work, we became aware of related works by R. Hussein {\it et al.}~\cite{hussein_nonlocal_2018}, and by N. S. Kirsanov {\it et al.}~\cite{kirnasov}.


\bibliographystyle{apsrev}

\end{document}